\documentclass[aps,prc,11pt,superscriptaddress,nofootinbib]{revtex4}

\usepackage[usenames]{color}
\usepackage[dvips]{graphicx}
\usepackage{amsmath}
\usepackage{amsfonts}
\usepackage{amssymb}
\usepackage{mathrsfs}
\usepackage{bm}
\usepackage{verbatim}
\usepackage{cancel}

\newcommand{\beq}{\begin{equation}}
\newcommand{\eeq}{\end{equation}}
\newcommand{\bea}{\begin{eqnarray}}
\newcommand{\eea}{\end{eqnarray}}

\newcommand{\mlo}{M_{\text{lo}}}
\newcommand{\mhi}{M_{\text{hi}}}
\newcommand{\chn}[3]{{{}^{#1}\text{#2}_{#3}}}
\newcommand{\cs}[2]{\chn{#1}{S}{#2}}
\newcommand{\cp}[2]{\chn{#1}{P}{#2}}
\newcommand{\cd}[2]{\chn{#1}{D}{#2}}
\newcommand{\cf}[2]{\chn{#1}{F}{#2}}
\newcommand{\csd}{{\cs{3}{1}-\cd{3}{1}}}
\newcommand{\cpf}{{\cp{3}{2}-\cf{3}{2}}}
\newcommand{\mnda}{\overline{\text{NDA}}}

\preprint{JLAB-THY-11-1401}
\preprint{INT-PUB-11-035}

\begin{document}

\title{Renormalizing chiral nuclear forces: A case study of $\cp{3}{0}$}
\author{Bingwei Long}
\email{bingwei@jlab.org}
\affiliation{Excited Baryon Analysis Center (EBAC), Jefferson Laboratory, 12000 Jefferson
Avenue, Newport News, Virginia 23606, USA}
\author{C.-J. Yang}
\email{cjyang@email.arizona.edu}
\affiliation{Department of Physics, University of Arizona, Tucson, Arizona 85721, USA}
\affiliation{Institute of Nuclear and Particle Physics, Ohio University, Athens, Ohio
45701, USA}
\date{\today }

\begin{abstract}
We discuss in this paper the subleading contact interactions, or counterterms, in the $\cp{3}{0}$ channel of nucleon-nucleon scattering up to $O(Q^{3})$, where, already at leading order, Weinberg's original power counting (WPC) scheme fails to fulfill renormalization group invariance due to the singular attraction of one-pion exchange. Treating the subleading interactions as perturbations and using renormalization group invariance as the
criterion, we investigate whether WPC, although
missing the leading order, could prescribe correct subleading counterterms.
We find that the answer is negative and, instead, that the structure of counterterms agrees with a modified version of naive dimensional analysis. Using $\cp{3}{0}$ as an example, we also study the cutoffs where the subleading potential can be iterated together with the leading one.
\end{abstract}
\maketitle

Since its proposal~\cite{Weinberg:1990-1991} and first implementation~\cite{Ordonez:1993-1995}, the effective field theory (EFT) description of nuclear
forces has enjoyed significant phenomenological successes~\cite{Ordonez:1993-1995, Epelbaum, idaho, Epelbaummore, reviews}. Based on naive dimensional analysis (NDA), Weinberg's original power counting (WPC) scheme has one crucial assumption: each derivative on the Lagrangian terms is always suppressed by the underlying scale of chiral EFT, $M_{\text{hi}}$, where $M_{\text{hi}}\sim m_{\sigma}$ with $m_{\sigma}$ the mass of the $\sigma$ meson. Due to the lack of bound states or resonances near threshold (with the delta-isobar integrated out), this assumption turns out to be quite applicable in the single-nucleon sector. However, the nonperturbative nature of few-nucleon systems could make infrared mass scale ($M_{\text{lo}}\sim 100$ MeV)---such as the pion decay constant $f_{\pi} \simeq 92$ MeV and the pion mass $m_{\pi}\simeq 140$ MeV---to enhance $NN$ contact interactions relative to WPC. In fact many works have pointed out that WPC is inconsistent from the point of view of renormalization group (RG) invariance~\cite{wpciswrong, Nogga:2005hy, Birse:2009my, Valderrama:2009ei, Valderrama:2011mv}. Here renormalization is associated with ultraviolet (UV) momenta in the Lippmann-Schwinger (LS) equation rather than with the UV divergence of perturbative, $NN$ irreducible diagrams.

One mechanism to upset WPC is the singular (diverging at least as fast as $1/r^2$) attraction of the tensor force of one-pion exchange (OPE), $-1/r^3$ at $r \to 
0$, in certain S, P, and D channels where OPE needs full iteration: $\csd$, $\cp{3}{0}$, $\cpf$, and $\cd{3}{2}$. The singular attraction of OPE requires a counterterm at leading order (LO) in these channels. However, the P or D wave counterterms are counted as subleading in WPC since they have at least two derivatives~\cite{Nogga:2005hy}. In other words, the LO counterterm of, say, $\cp{3}{0}$, is enhanced by $\mathcal{O}(\mhi^2/\mlo^2)$. We denote different orders of the EFT expansion by its relative correction to the LO. Thus, the next-to-leading order (NLO) is labeled by $\mathcal{O}(Q/M_{\text{hi}})$ or $\mathcal{O}(Q)$ for short, and next-to-next-to-leading order (NNLO) by $\mathcal{O}(Q^{2}/M_{\text{hi}}^{2})$ or $\mathcal{O}(Q^2)$, and so on.

The question remains as to how to modify WPC at subleading orders. The model study in Ref.~\cite{Long:2007vp} suggests, referred to in the paper as modified NDA ($\mnda$), that in the case of the LO long-range potential being singular and attractive, the subleading counterterms are enhanced relative to NDA by the same amount as the LO counterterms. As a first study, we use $\cp{3}{0}$ in the paper to investigate whether $\mnda$ is applicable in nuclear EFT. Note that different approaches toward renormalization of nuclear forces are offered in Refs.~\cite{alternative, saopaulo, spaniards, yanglo, Yang:2009kx, Yang:2009pn}, and that perturbative pion theory leads to a very different renormalization program~\cite{pertpion}.

The resummation of ladder diagrams in lower partial waves can be achieved by solving the LS equation, with the schematic form,
\begin{equation}
T= V + \int^{\Lambda} V G T \, ,
\label{eqn:LSE-PW}
\end{equation}
where the potential $V$ consists of short-range counterterms $V_{S}$ and
long-range pion-exchange $V_{L}$, $G$ is the Schr\"{o}dinger propagator and 
$\Lambda$ is a (sharp) momentum cutoff. Being singular and attractive in $\cp{3}{0}$,
OPE demands promotion of the leading $\cp{3}{0}$ counterterm,
\begin{equation}
\langle \,{\cp{3}{0}}|V_{S}^{(0)}|\,{\cp{3}{0}}\rangle = C_{\cp{3}{0}}\,p^{\prime}p\sim \frac{4\pi }{m_{N}}\frac{p' p}{M_{\text{lo}}^{3}} \, ,
\end{equation} 
an $\mathcal{O}(\mhi^2/\mlo^2)$ enhancement over WPC~\cite{Nogga:2005hy}, where $4\pi/m_N$ is a common factor associated with nucleonic loops. For a comparison, the leading S-wave counterterms scale generically as $C_\text{S} \sim 4\pi/(m_N\mlo)$.

It will prove useful to have the short-range behavior of the LO $\cp{3}{0}$ wave function. For $kr\ll 1<\Lambda r$ where $k$ is the center-of-mass momentum, the LO
wave function $\psi _{k}^{(0)}(r)$ can be solved for in powers of $k^{2}$~\cite{frank,PavonValderrama:2007nu}, up to
a normalization factor,
\begin{equation}
\psi_{k}^{(0)}(r) \sim \left(\frac{\lambda}{r}\right) ^{\frac{1}{4}}\left[
u_{0} + k^2r^2 \sqrt{\frac{r}{\lambda}}u_{1} +
\mathcal{O}(k^{4}) \right] \,,  \label{eqn:psi0}
\end{equation}
where $\lambda =\frac{3g_{A}^{2}m_{N}}{32\pi f_{\pi }^{2}}$ with $g_A = 1.29$ the nucleon axial charge, $u_{0}$ and $
u_{1}$ are oscillatory functions in terms of $r/\lambda$ and $\phi$ with amplitudes $\sim 1$, where $\phi$ is the phase between the two independent solutions and is related to $C_\cp{3}{0}$.

Before two-pion exchanges (TPEs) are accounted for, which provide $\mathcal{O}(Q^2)$ or higher corrections to OPE, there might exist a nonvanishing $\mathcal{O}(Q)$ manufactured by inserting subleading
contact interactions alone. In fact, Ref.~\cite{Birse:2009my} rated the
triplet P-wave subleading contact interactions as $\mathcal{O}(Q^{3/2})$,
a lower-order contribution than TPE. 

We argue that a nonvanishing $\mathcal{O}(Q)$ does not seem to be necessary.
The study of Ref.~\cite{PavonValderrama:2007nu} suggests that the residual
cutoff dependence of the renormalized LO $\cp{3}{0}$ amplitude is $\mathcal{O}(\Lambda ^{-5/2})$. Had this residual cutoff dependence vanished
slower than $\mathcal{O}(\Lambda^{-2})$, our ignorance of short-range
physics would have been larger than $\mathcal{O}(Q^{2}/M_{\text{hi}}^{2})$,
that is, larger than what TPE could compensate. 
If this happened, one would have had to consider an $\mathcal{O}(Q/M_{\text{hi}})$
correction to the LO amplitude induced by inserting a four-derivative $\cp{3}{0}$ counterterm, before accounting for TPE. However,
the rather small residual cutoff dependence of the LO $\cp{3}{0}$ amplitude does not
ask for a nonvanishing $\mathcal{O}(Q)$.

With $\mathcal{O}(Q)$ vanishing, the $\mathcal{O}(Q^{2})$ and $\mathcal{O}
(Q^{3})$ amplitudes, $T^{(2)}$ and $T^{(3)}$,
consist, respectively, in one insertion of $\mathcal{O}(Q^{2})$ or $\mathcal{O}
(Q^{3})$ potentials,
\begin{equation}
\begin{split}
T^{(2,\, 3)}& =V^{(2,\, 3)}+\int^{\Lambda }V^{(2,\, 3)}GT^{(0)}+\int^{\Lambda
}T^{(0)}GV^{(2,\, 3)} \\
& + \int^{\Lambda}\int^{\Lambda}T^{(0)}G\,V^{(2,\, 3)}\,GT^{(0)}\, .
\label{eqn:LSE23}
\end{split}
\end{equation}
Since this is equivalent to first-order distorted wave expansion,
one can evaluate the ``superficial'' divergence of one insertion of $V_L^{(2)}$ (leading TPE) and $V_L^{(3)}$ (subleading TPE)
before any counterterm is considered, by investigating the short-distance behavior of the matrix element of $V_L^{(2,\, 3)}$ between the LO wave functions. This is facilitated by the short-distance behavior of the LO wave function~\eqref{eqn:psi0} and TPEs, $V_L^{(2)}\sim 1/r^{5}$ and $V_L^{(3)}\sim 1/r^{6}$.
With a radial coordinate cutoff $\mathcal{R}\sim 1/\Lambda $, we arrive at
\begin{align}
T^{(2)}& =\langle \psi ^{(0)}|V_L^{(2)}|\psi ^{(0)}\rangle \sim
\int_{\sim 1/\Lambda }drr^{2}|\psi ^{(0)}(r)|^{2}\frac{1}{r^{5}}  \notag \\
& \sim \alpha _{0}(\Lambda )\Lambda ^{5/2}+\beta
_{0}(\Lambda )k^{2}+\mathcal{O}(k^{4}\Lambda ^{-5/2}) ,
\label{eqn:supT2} \\
T^{(3)}& =\langle \psi ^{(0)}|V_L^{(3)}|\psi ^{(0)}\rangle \sim
\int_{\sim 1/\Lambda }drr^{2}|\psi ^{(0)}(r)|^{2}\frac{1}{r^{6}}  \notag \\
& \sim \alpha _{1}(\Lambda )\Lambda ^{7/2}+\beta
_{1}(\Lambda )\Lambda k^{2}+\mathcal{O}(k^{4}\Lambda ^{-3/2}) \, ,
\label{eqn:supT3}
\end{align}
where $\alpha_{0,1}(\Lambda)$ and $\beta_{0,1}(\Lambda )$ are oscillatory
functions diverging slower than $\Lambda$.
Their exact forms can be evaluated \cite{lucas} but are not crucial for our discussion.

It is not necessarily true that one must use two counterterms to subtract the two divergent terms
in Eq.~\eqref{eqn:supT2} or \eqref{eqn:supT3}. 
In fact, WPC prescribes only one counterterm up to $\mathcal{O}(Q^{3})$, and with the nonperturbative treatment it did
provide a good fit to partial wave analysis (PWA) for a moderate range of cutoffs \cite{Epelbaum, idaho, Epelbaummore}. We first consider the counterterms
prescribed by WPC,
\begin{equation}
\begin{split}
\langle {\cp{3}{0}}|V_{S}^{(2,\, 3)}|{\cp{3}{0}}\rangle =C_{{\cp{3}{0}}}^{(2,\, 3)}\,p^{\prime }p\,,
\end{split}
\label{eqn:3p0ccpresI}
\end{equation}
where we split $C_{{\cp{3}{0}}}$ into three pieces with $C_\cp{3}{0}^{(0)}$ determined at LO.
This splitting reflects the fact that the value of the ``bare'' $C_\cp{3}{0}$ could be modified at each order by, e.g., the short-range
core of TPE, but the number of physical, short-range inputs is still one.

The other scenario is to provide an equal number of counterterms as the
divergent terms in Eqs.~\eqref{eqn:supT2} and \eqref{eqn:supT3}:
\begin{equation}
\begin{split}
\langle {\cp{3}{0}}|V_{S}^{(2,\, 3)}|{\cp{3}{0}}\rangle = C_{{\cp{3}{0}}
}^{(2,\, 3)}\,p^{\prime }p+D_{{\cp{3}{0}}}^{(0,\, 1)}\,p^{\prime }p({p^{\prime}}
^{2}+p^{2})\,.
\end{split}
\label{eqn:3p0ccpresII}
\end{equation}
Thus, up to $\mathcal{O}(Q^{3})$, every $\cp{3}{0}$
counterterm gets enhanced relative to WPC by the same amount, $\mathcal{O}(M_{\text{hi}}^{2}/M_{\text{lo}}^{2})$. This is exactly what to be expected based on $\mnda$~\cite{Long:2007vp}.

There are a few versions of TPEs in the literature with slight differences in how double counting is avoided~\cite{Ordonez:1993-1995, Epelbaum, idaho, TPEs}.
For definiteness, we use the version in Ref.~\cite{Epelbaum},
delta-less TPE expressions with dimensional regularization. We adopt the
following low-energy constants for the $\nu = 1$ $\pi \pi NN$ seagull couplings (GeV$^{-1}$): 
$c_{1}=-0.81$, $c_{3}=-4.7$ and $c_{4}=3.4$~\cite{Epelbaum}.

We first test the prescription given by WPC~\eqref{eqn:3p0ccpresI}. At each
order, $C_{{\cp{3}{0}}}$ is determined such that the phase shift at $T_{\text{lab}}=50$ MeV agrees with the Nijmegen PWA~\cite{Stoks:1993tb}. Note that
extra cares are needed to convert the subleading $T$-matrix into phase shifts, as shown in the Appendix.

Figure~\ref{fig:fxk-pres1} shows the resulting
phase shifts as functions of $\Lambda$ at $T_{\text{lab}}=40$, $80$, and $130$ MeV.
At $\mathcal{O}(Q^{2})$, the oscillatory cutoff dependence becomes more evident as the energy increases.
This is consistent with the superficial divergence \eqref{eqn:supT2}; while
the $\alpha_{0}$ term is taken care of by $C_{\cp{3}{0}}^{(2)}$, the
amplitude of the oscillation of $\beta_{0}(\Lambda)k^{2}$ is left
intact and grows as the energy increases. Moving on to $\mathcal{O}(Q^{3})$
we find more drastic, oscillatory cutoff
dependence with a visibly growing amplitude. This is due to the factor of $\Lambda$ that accompanies the oscillatory $\beta_{1}(\Lambda)$ in Eq.~\eqref{eqn:supT3}.

\begin{figure}
\centering
\begin{tabular}{rrr}
\includegraphics[scale=0.5, clip=true]{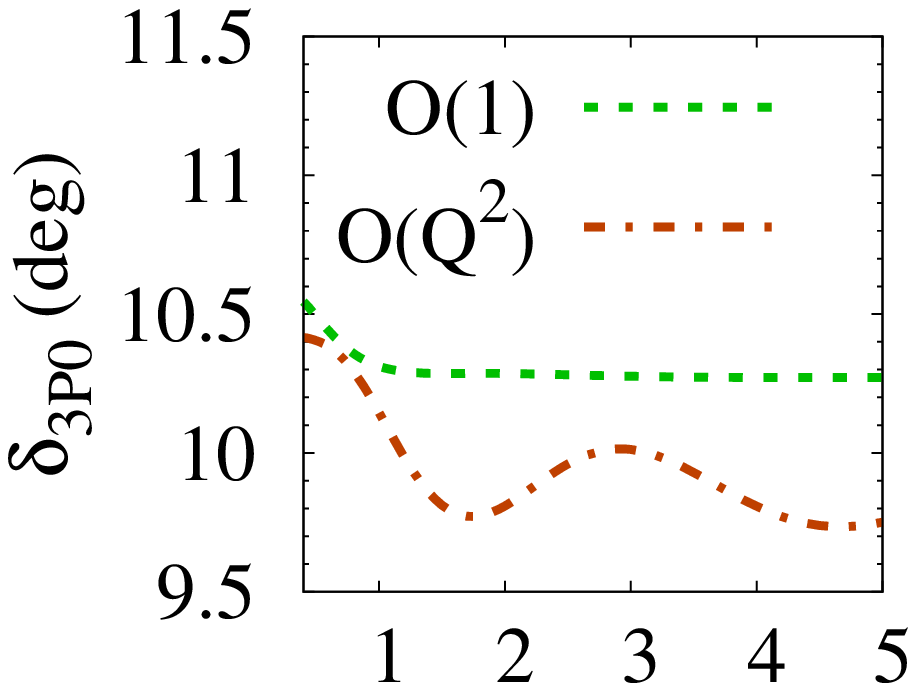} &
\includegraphics[scale=0.5, clip=true]{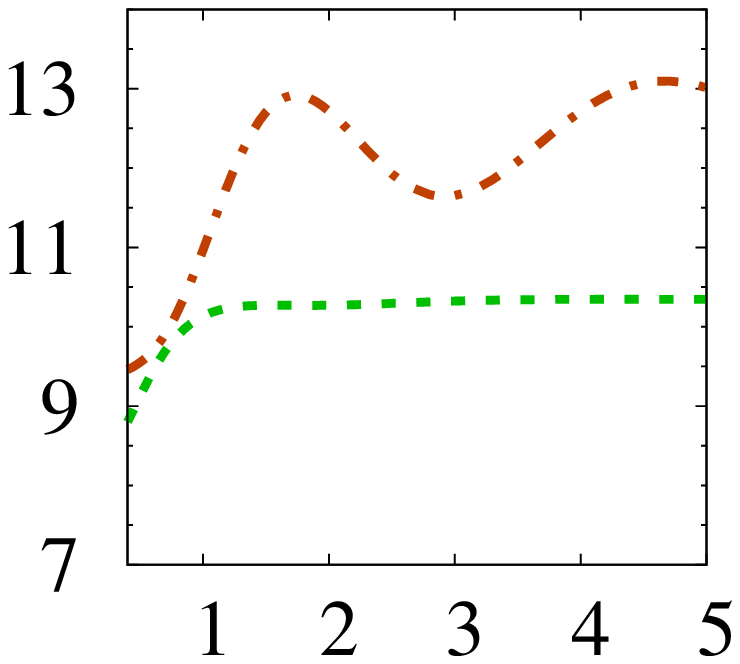} &
\includegraphics[scale=0.5, clip=true]{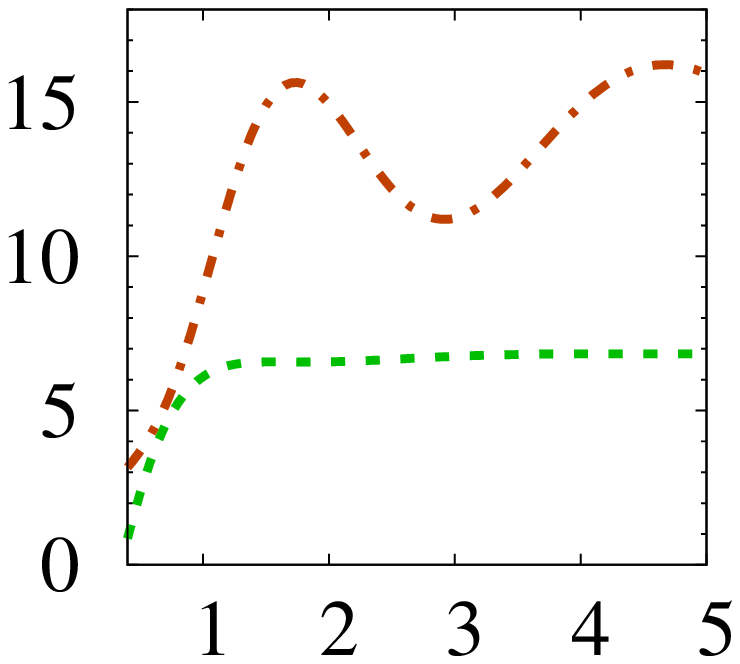} \\
\includegraphics[scale=0.5, clip=true]{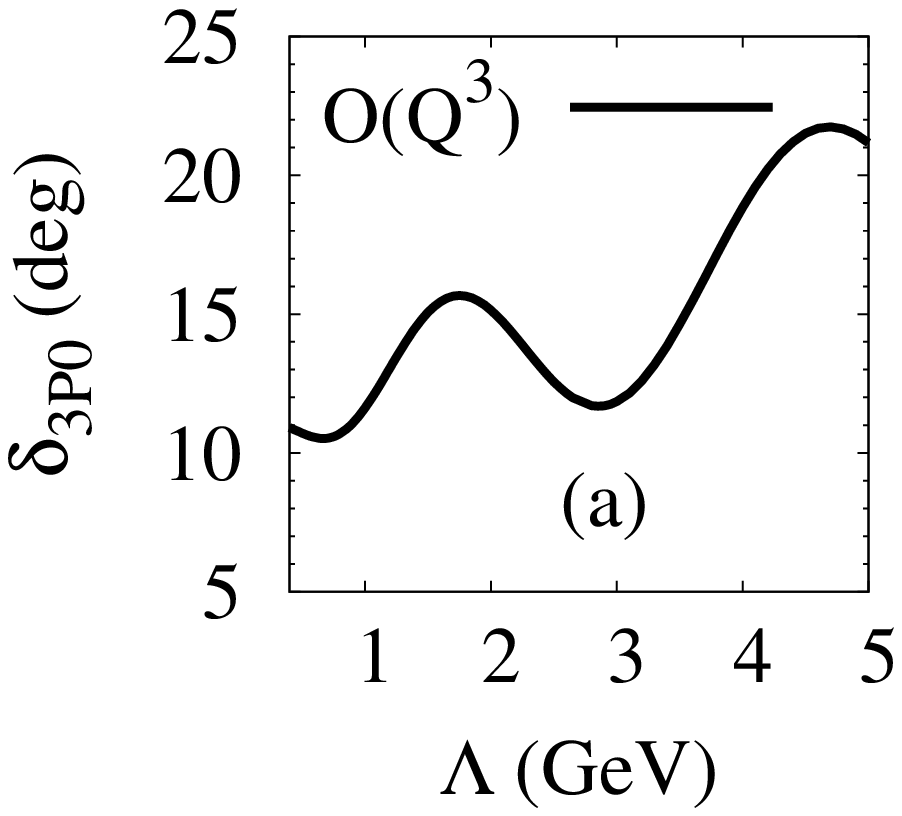} &
\includegraphics[scale=0.5, clip=true]{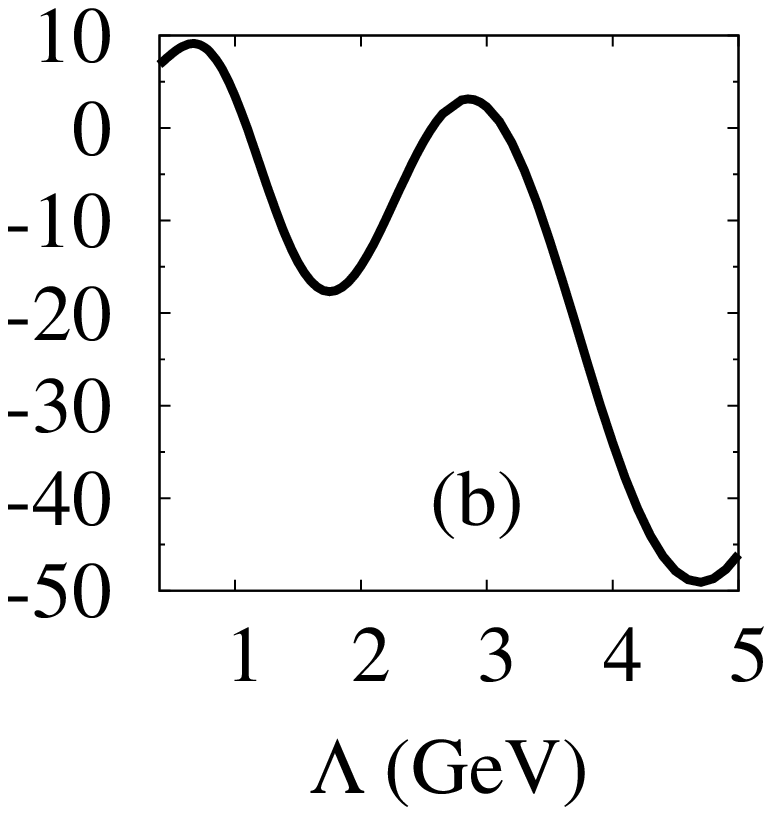} &
\includegraphics[scale=0.5, clip=true]{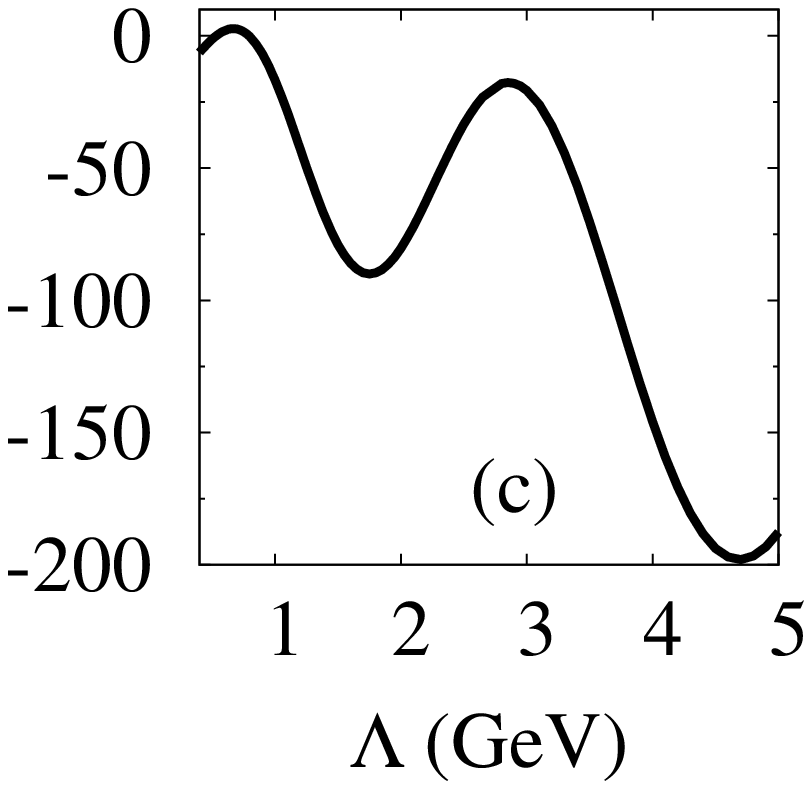}
\end{tabular}
\centering
\caption{(Color online) With the subleading counterterms \eqref{eqn:3p0ccpresI}, the $\mathcal{O}(1)$, $\mathcal{O}(Q^{2})$ (upper row) and $\mathcal{O}(Q^{3})$ (lower row) ${\cp{3}{0}}$ phase shifts as
functions of the momentum cutoff at $T_{\text{lab}}$ = $40$ (a), $80$
(b), and $130$ (c) MeV.}
\label{fig:fxk-pres1}
\end{figure}

\begin{figure}
\centering
\includegraphics[scale=0.5, clip = true]{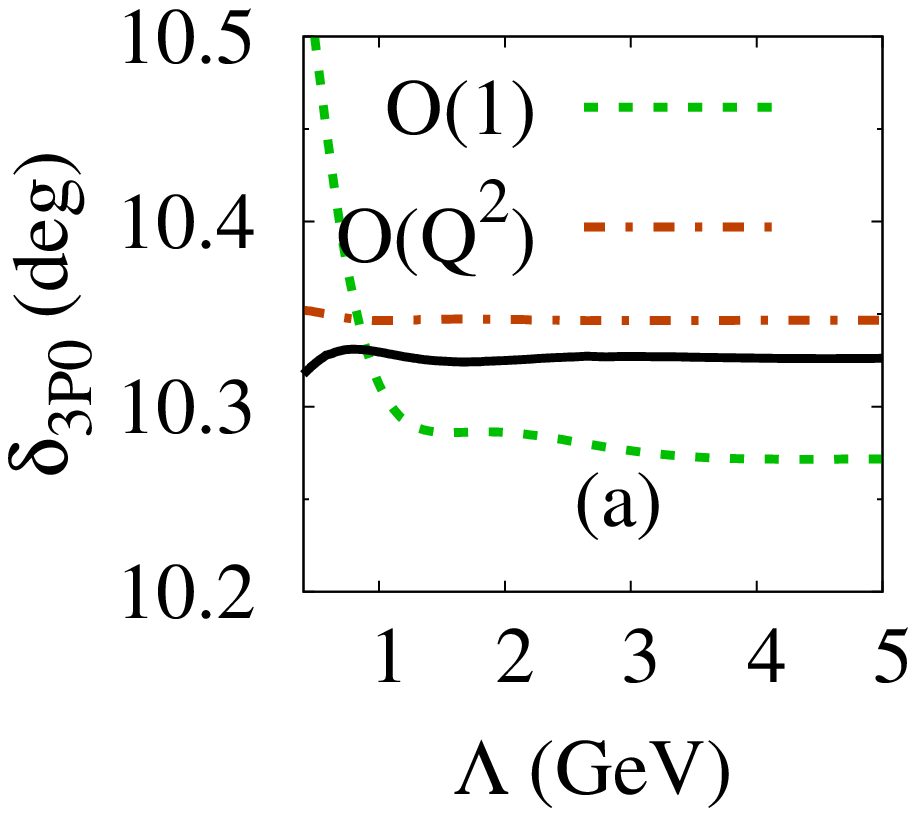}
\includegraphics[scale=0.5, clip = true]{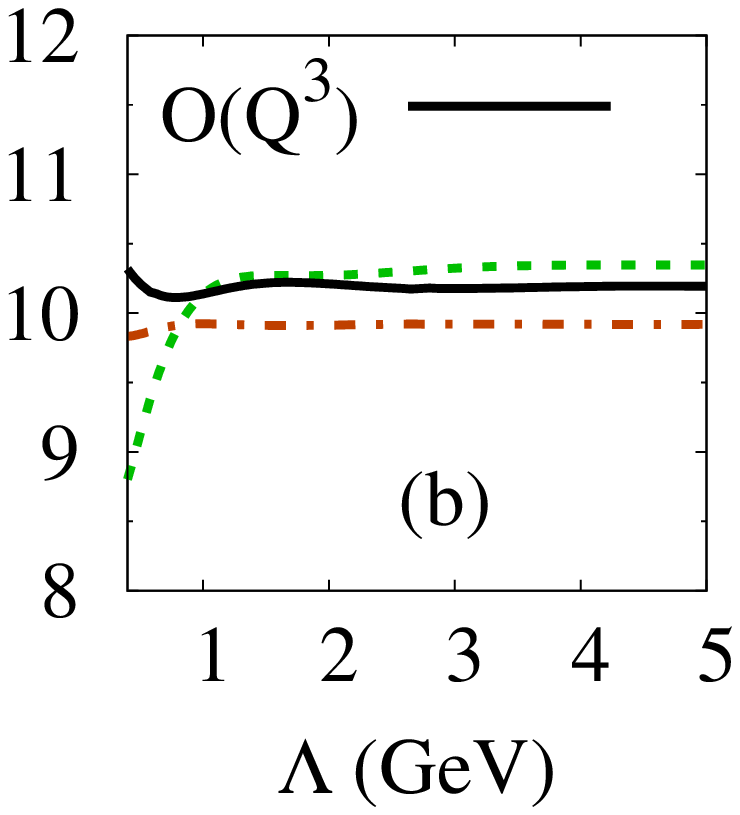}
\includegraphics[scale=0.5, clip = true]{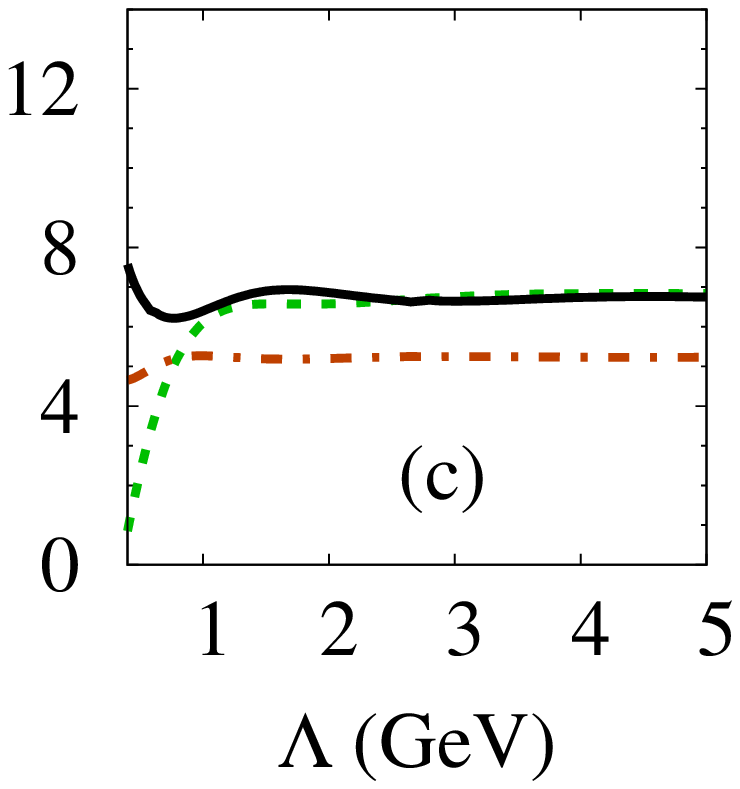}
\centering
\caption{(Color online) With the subleading counterterms \eqref{eqn:3p0ccpresII}, the $\mathcal{O}(1)$, $\mathcal{O}(Q^2)$, and $\mathcal{O}(Q^3)$ $\cp{3}{0}$ EFT
phase shifts as functions of the momentum cutoff at $T_\text{lab} = 40$
(a), $80$ (b), and $130$ (c) MeV.}
\label{fig:fxk-pres2}
\end{figure}

With the $\mnda$ counting \eqref{eqn:3p0ccpresII}, we need two physical inputs
to determine the values of $C_\cp{3}{0}$ and $D_\cp{3}{0}$. We fit them to reproduce the PWA at $T_\text{lab} = 20$ and $50$ MeV. The plateaus in Fig.~\ref{fig:fxk-pres2} clearly show the RG invariance of the power counting \eqref{eqn:3p0ccpresII}, where the phase shifts are plotted as functions of $\Lambda$ at given $T_{\text{lab}}$.

\begin{figure}
\centering
\includegraphics[scale=0.32, clip=true]{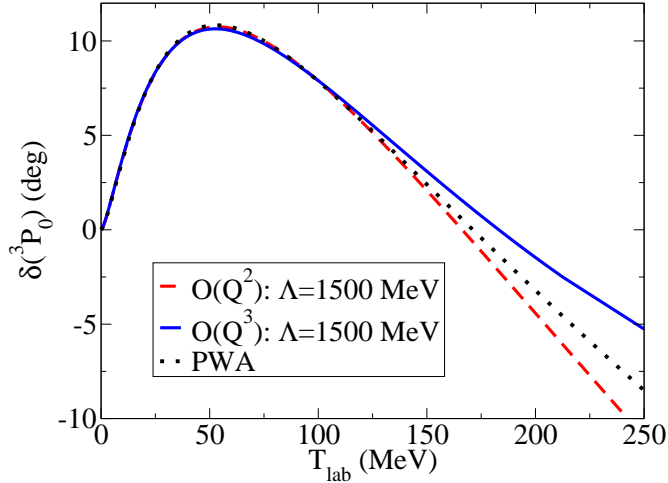} \centering
\caption{(Color online) With the subleading counterterms \eqref{eqn:3p0ccpresII} and the
improved fitting procedure, the ${\cp{3}{0}}$ EFT phase shifts at $%
\mathcal{O}(Q^{2})$ and $\mathcal{O}(Q^{3})$ as function of $T_\text{lab}$
for $\Lambda =1500$ MeV.}
\label{fig:phs3p0-updated}
\end{figure}

In Fig.~\ref{fig:phs3p0-updated}, the EFT phase shifts are plotted as function of energy. The fit is refined by employing 
more PWA points ($T_{\text{lab}}=$25, 50, 75, and 100 MeV) in the fitting procedure.
We see that both $\mathcal{O}(Q^{2})$ and $\mathcal{O}(Q^{3})$ are in good agreement with the PWA.

Although the perturbative treatment of WPC does not lead to cutoff independent
results,
the nonperturbative treatment does seem to fulfill RG invariance~\cite{Yang:2009kx}. 
It is therefore instructive to compare the
following three scenarios for $\cp{3}{0}$: (i) the perturbative (Pert-CD) and (ii) nonperturbative calculations with the modified power counting \eqref{eqn:3p0ccpresII} (Iter-CD), and (iii) the nonperturbative calculation with WPC (Iter-WPC).

\begin{figure}
\centering
\includegraphics[scale=0.32]{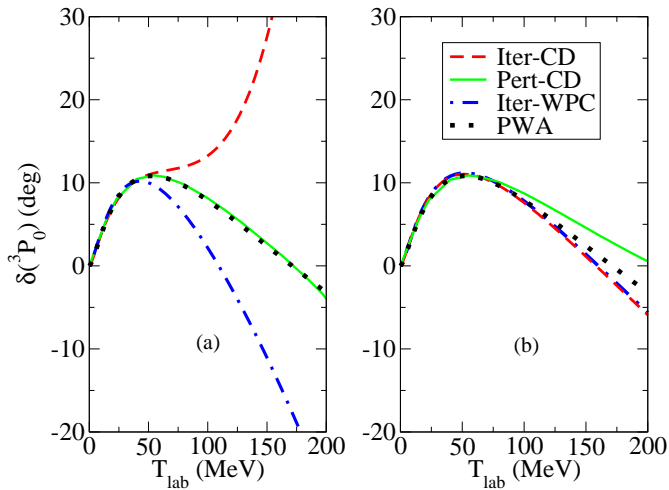}
\caption{(Color online) The $\cp{3}{0}$ phase shifts by Iter-CD, Pert-CD and Iter-WPC
as functions of laboratory energy. (a) is plotted with $\Lambda = 400$
MeV and (b) with $\Lambda = 1200$ MeV.}
\label{fig:window}
\end{figure}

Shown in Fig.~\ref{fig:window} are the $\cp{3}{0}$ phase shifts calculated
at $\mathcal{O}(Q^{3})$ with the aforementioned three schemes, where
the fit of $C_\cp{3}{0}$ and $D_\cp{3}{0}$ in Eq.~\eqref{eqn:3p0ccpresII} is performed with the PWA inputs up to $T_{\text{lab}}=50$ MeV. 
At the lower end of cutoffs (as exemplified by $\Lambda = 400$ MeV), the three curves differ drastically from each other above $T_{\text{lab}} = 50$ MeV, with the Pert-CD curve agreeing somewhat better with the PWA. As the cutoff goes higher (exemplified by $\Lambda =1200$ MeV),
the difference between Iter-CD and Iter-WPC becomes smaller and eventually
vanishes, in accordance with the finding of Ref.~\cite{En08}. We note that this does not necessarily mean that the fitting drives $D_{^3\!P_0}$ to 0. Rather, the quality of the fit is not sensitive to $D_\cp{3}{0}\Lambda^2/C_\cp{3}{0}$ when the ratio is tuned from 0 to 1.

In summary, we conclude:

\begin{itemize}
\item[(i)] WPC does not accommodate a cutoff independent $T$-matrix at $\mathcal{O}(Q^{2})$ or $\mathcal{O}(Q^{3})$ when subleading potentials are treated as perturbations on top of the LO.

\item[(ii)] RG invariance can be achieved by the modified power counting \eqref{eqn:3p0ccpresII}, based on modified naive dimensional analysis. This suggests that
$-1/r^2$, the LO long-range potential in the model of Ref.~\cite{Long:2007vp}, is not crucial for $\mnda$ to be applicable.

\item[(iii)] At $\mathcal{O}(Q^3)$, Iter-CD, Pert-CD and Iter-WPC show that in a limited range of cutoffs these three approaches produce similar phase shifts for $\cp{3}{0}$, a conclusion similar to that of Refs.~\cite{Yang:2009kx, Yang:2009pn}. While it is instructive, we refrain from drawing the same conclusion for other channels; it may well be that the ``common'' window of cutoffs appears at different location for different channels.
\end{itemize}

A simultaneous, coordinate-space calculation in Ref.~\cite{Valderrama:2011mv} has come to our attention. The conclusion drawn there for $\cp{3}{0}$ agrees with ours, that is, in agreement with $\mnda$. However, Refs.~\cite{Valderrama:2009ei, Valderrama:2011mv} concluded a proliferation of six counterterms in each of the coupled channels, $\csd$ and $\cpf$, whereas $\mnda$ suggests three once WPC is corrected at LO. We defer to a further momentum-space calculation of the triplet channels~\cite{BwLCJYtriplet} that are subject to the singular attraction of OPE to investigate whether $\mnda$ or the conclusion of Refs.~\cite{Valderrama:2009ei, Valderrama:2011mv} on the coupled channels can be verified.

We thank U.~van Kolck, D.~Phillips, E.~Ruiz Arriola, and E.~Epelbaum for useful discussions. CJY thanks B.~Barrett for valuable
support. We are grateful for hospitality to the National Institute for Nuclear Theory (INT) at the University of Washington and the organizers of the INT program ``Simulations and Symmetries: Cold Atoms, QCD, and Few-hadron Systems'', at which the work was initiated. This work is supported by the US DOE under Contract Nos. DE-AC05-06OR23177 (BwL), DE-FG02-04ER41338 (CJY) and by the NSF under Grant No. PHYS-0854912 (CJY). This work is coauthored by Jefferson Science Associates, LLC under U.S. DOE Contract No. DE-AC05-06OR23177.

\appendix
\section{\label{sec:pertconv}}
In distorted wave expansion, the unitarity of the $S$-matrix no longer
rigorously holds. Therefore, the normalization of the $T$-matrix we adopt in the paper,
$T=-e^{i\delta}\sin \delta\,/\, m_{N}k$,
is no longer exact at subleading orders. Suppose that, with $T^{(1)}$ vanishing, the $T$-matrix
and the phase shifts have the following expansion: $T=T^{(0)}+T^{(2)}+T^{(3)}+\cdots$, $\delta =\delta ^{(0)}+\delta
^{(2)}+\delta ^{(3)}+\cdots $.
Treating $%
T^{(2),\, (3)}$ and $\delta ^{(2),\, (3)}$ as perturbations,
one finds
\begin{equation}
T^{(2),\,(3)}=-\delta ^{(2),\,(3)}\frac{e^{2i\delta ^{(0)}}}{m_{N}k}\,.
\end{equation}

\end{document}